# Available Transfer Capability Calculation for Wind-Integrated Power Systems Considering Wind Speed Spatiotemporal Correlation and Primal-Dual Interior Point Method

HUANGPU Xia-Liang

(School of Electrical Engineering, North China University of Water Resources and Electric Power, Zhengzhou City, Henan Province 450046, China)

**Abstract**: This paper explores the intricate effects of wind power integration on the Available Transfer Capability (ATC) of power systems, emphasizing the significance of spatiotemporal correlations in wind speed. We present an innovative optimal power flow model that integrates the Primal-Dual Interior Point Method (PDIPM), ensuring both computational accuracy and efficiency. This research pioneers a systematic analysis of how spatiotemporal wind speed correlations influence wind power output, thereby refining ATC calculations and improving prediction reliability. Furthermore, we assess the impacts of wind farm integration capacity, location, and connection methods on ATC, offering valuable insights for power system planning and market operations.

## 1 Introduction

Wind power bases in China are primarily constructed in the "Three North" regions. These areas have low load levels, weak grid structures, and are distant from load centers, necessitating large-scale, long-distance power transmission [1]. This specific distribution of load and wind resources dictates that wind power development must adopt the model of "constructing large bases and integrating into a large grid" [2-3]. With the rapid increase in the proportion of wind power, its randomness and volatility pose challenges to the safe and stable operation of the grid [4], making the calculation of the Available Transfer Capability (ATC) between regions more complex. Traditional transmission capability calculations could predict and meet peak load demands by controlling maximum generation power. However, large-scale integration of wind power increases the system's uncertainty and operational risks, reducing the grid's safety margin [5]. Therefore, in-depth research on the safety and stability of wind power integration, along with effective technical measures, is crucial for ensuring the stable operation of the power system and the smooth functioning of the electricity market.

In the 1970s, the concept of regional power exchange capability (transmission interchange capability or simultaneous interchange/transfer capability) began to receive attention in power systems [6]. Initially used as a reference for internal system scheduling, it wasn't until the U.S. Federal Energy Regulatory Commission (FERC) issued Order No. 889 in 1996, requiring grid owners to calculate and disclose ATC between regions, which the significance of ATC in the electricity market became increasingly prominent [7]. The North American Electric Reliability Council (NERC) defines ATC as the remaining transmission capacity in the grid above the existing electric energy transmission contracts [8]. The operational safety and reliability of power systems are deeply affected by their transmission capability, especially in the context of the electricity market environment, where system uncertainty and the complexity of energy transactions make ATC a critical component of system security [9].

In recent years, the study of ATC between regional power systems, due to its computational complexity, has attracted considerable attention. ATC is used to assess the residual transmission capacity over a given period on a transmission section, guiding the safe and stable operation of the power system. ATC calculations are divided into online and offline methods. Online ATC calculations usually employ deterministic algorithms, such as the linear distribution factor method, continuous power flow method, optimal power flow method, and AC sensitivity analysis method, to achieve rapid calculations and meet real-time demands [10-13]. Offline ATC calculations, due to longer prediction times and the influence of uncertainty factors, use probability-based uncertainty algorithms, such as enumeration methods, stochastic programming [14], Monte Carlo simulation [15], and robust planning methods [16], employing mathematical statistics to handle uncertain factors and improve calculation accuracy.

In recent years, as wind power has rapidly gained ground as a sustainable energy solution globally [17], the large-scale grid integration of wind power has posed new challenges to the calculation of ATC for power systems. Studies show that the spatiotemporal characteristics of wind speed significantly affect the accuracy of ATC predictions, such as large wind farm output power optimization based on Weibull distribution modeling in [18], wind speed time series modeling and its probabilistic assessment of ATC in [19], and the improved algorithms for continuous power flow methods in [20]. Additionally, reference [21] performed reduced-order analysis for system planning under wind power integration, considering load fluctuations. However, existing methods still show deficiencies in considering the temporal and spatial correlations of wind speed, which may lead to significant deviations between predicted ATC for wind-integrated systems and actual operating characteristics [22-23].

As renewable energy, particularly wind power, continues to develop rapidly and integrate on a large scale, power systems face unprecedented uncertainties, posing severe challenges to their safety and stable operation. Wind power is characterized by its volatility and intermittency [24], making grid load forecasting and management increasingly complex [25-26]. This complexity arises not only from the inherent instability of wind power but also because wind farms are typically located far from load centers, exacerbating the difficulty and uncertainty of power transmission. To address these challenges, grids must adopt more advanced technologies to enhance system adaptability and flexibility [27]. Additionally, the rapid increase in new types of loads, such as electric vehicles, also poses new demands on grid operation [28-30]. The increasing charging demands of electric vehicles raise the peak loads of grids, and the randomness of charging behavior makes forecasting and scheduling grid loads more difficult. The integration of these new types of loads not only alters the load patterns of grids but may also inadvertently cause fluctuations in grid frequency and voltage, affecting overall grid performance [31-33]. Moreover, as the digitalization and networking of power systems deepen, the issue of information security in grids has become increasingly prominent [34-36]. Digitalization of grid control systems, while improving operational efficiency, also exposes grids to threats from cyber-attacks [37]. These attacks can disrupt the normal operation of grids and even lead to large-scale power outages. Therefore, protecting grids from information security threats and ensuring the security of grid data and operating systems have become indispensable parts of grid design and operation [38-39]. In this context, accurately and efficiently calculating the grid's Available Transfer Capability (ATC) becomes especially critical. ATC is not only a key indicator of the grid's ability to safely transmit electric power but also a fundamental aspect of power market operations. Precise calculations of ATC can help grid operators better plan grid expansions, optimize generation resource allocation, and effectively address challenges posed by wind power integration and new types of loads. This requires grid operators to employ advanced simulation tools and algorithms, such as models considering the spatiotemporal correlation of wind speeds, to achieve more accurate predictions of wind power output and load fluctuations.

With the depletion of non-renewable resources and the intensifying environmental pollution from fossil fuel combustion, the development and utilization of renewable energy have rapidly advanced. Wind power, as an economically beneficial, clean, and pollution-free green renewable energy technology, has been widely applied globally [40]. Due to the increasing scale and transmission capacity of wind power, its randomness, intermittency, and the spatiotemporal correlation increasingly impact the output power of wind farms. Therefore, analyzing the operational characteristics of power systems after wind power integration has garnered attention, and ATC is an important indicator for assessing system operational stability. Accurately and efficiently calculating ATC is crucial for ensuring the safe and stable operation of power systems and smooth transactions in the power market [41-42].

According to the definition of ATC provided by the North American Electric Reliability Council (NERC), the calculation of ATC requires quantifying the impact of various uncertainties, including generation scheduling uncertainty, load fluctuations, and component failures. As large-scale wind power is integrated into the grid, the uncertainty of wind power output becomes another significant factor. Current domestic and international research on ATC includes the following: References [43-44] consider the impact of system faults and load fluctuations on ATC but do not take into account the impact of wind power integration. As China's installed wind power capacity ranks first in the world, and wind power occupies an increasingly significant proportion of the entire power generation system, accurately assessing the system's ATC must consider the influence of wind speed and other uncertainty factors. To study the impact of wind power integration on system ATC, some domestic researchers have investigated ATC situations after wind power integration through measures such as the probability distribution of wind speeds, time series models of wind speeds, sensitivity of power fluctuations at wind power nodes, and scenario aggregation based on wind power levels. Although these studies have considered the intermittency and variability of wind speeds on ATC, the fact is that wind speed distributions across different wind farms have strong correlations. The correlation of wind speeds directly affects the output power of wind farms and further influences the fluctuation trends of ATC. Therefore, to more accurately analyze the impact of wind power integration on system ATC, wind speed correlation should be considered in wind speed predictions; Reference [45] uses a semi-invariant method to solve for dynamic stochastic power flow considering correlations, and Reference [46] proposes a new real-time wind speed prediction method based on spatiotemporal correlation and BP neural networks to model wind speed correlation. However, these studies only generally consider wind speed correlation. Wind speeds not only have temporal correlation but also spatial correlation, and the spatiotemporal correlation of wind speeds has a certain impact on prediction errors. This paper considers the temporal and spatial correlations of wind speeds in detail, resulting in wind speed predictions that more accurately reflect the actual wind speeds of wind farms, making the analysis results of system ATC more reliable and more in line with reality.

In response to the aforementioned issues, the method of spatiotemporal correlation prediction error proposed by Christopher Scott Saunders is used to address the issue of spatiotemporal correlation of wind speeds, considering the spatiotemporal correlation of wind speeds to make the ATC analysis results more accurate. A spatial correlation matrix is constructed and decomposed through singular value decomposition. The decomposed unitary and diagonal matrices are applied to the wind speed prediction error matrix, resulting in a wind speed prediction error matrix with spatial correlation. Through time autocorrelation matrices and variance matrices, time covariance matrices are constructed, and the same method yields unitary and diagonal matrices. The spatial correlation wind speed prediction error matrix is modified, resulting in a spatiotemporal correlation wind speed prediction error, which is superimposed with the wind speed predictions to obtain wind speed predictions with spatiotemporal correlation. From the wind power curve, wind power output with spatiotemporal correlation is obtained [47].

This paper considers the impact of the spatiotemporal characteristics of wind speeds and load fluctuations on the system, proposing an ATC calculation model under optimal power flow (OPF), and optimizes the modified model using the primal-dual interior point method (PDIPM). Different scenarios are compared to analyze the effects of different wind speed correlation coefficients, wind power integration capacity, wind power integration locations, and wind power integration methods on system ATC, providing reference for wind farm planning and power market operations.

## 2   Wind Farm Model

## 2.1 Wind Speed Model Considering Spatiotemporal Correlation

The actual wind speed can be considered as a superposition of the predicted wind speed and the prediction error, which can be described by the following formula:

$$v(\omega,h) = F(\omega,h) + W_{ST}(\omega,h) \quad \omega = 1,2,\cdots,N_W,\ h = 1,2,\cdots,N_H \tag{1}$$

Here, $v(\omega, h)$ represents the matrix of wind speed values with spatiotemporal correlation; $F(\omega, h)$ represents the matrix of predicted wind speed values; $W_{ST}(\omega, h)$ denotes the matrix of wind speed prediction errors with spatiotemporal correlation; $N_W$ stands for the number of wind fields; $N_H$ is the number of hours for which predictions are made。

Initially, wind speed predictions can be obtained through statistical forecasting methods found in reference [48]; then, using a matrix $W_h$ of independently and identically distributed Gaussian random variables to represent uncorrelated wind speed prediction errors [49], a spatiotemporal covariance matrix is constructed to modify $W_h$ to obtain wind speed prediction errors with spatiotemporal correlation [50].

To introduce spatial correlations between different wind fields, an $N_W \times N_W$ dimensional spatial correlation matrix $R_{S,h}$ is constructed:

$$R_{S,h} = \begin{bmatrix} \rho_{s,11} & \rho_{s,12} & \cdots & \rho_{s,1N_W} \\ \rho_{s,21} & \rho_{s,22} & \cdots & \cdots \\ \cdots & \cdots & \rho_{s,mn} & \cdots \\ \rho_{s,N_W1} & \cdots & \cdots & \rho_{s,N_WN_W} \end{bmatrix} \tag{2}$$

$$h = 1, 2, \cdots, N_H$$

In the formula, $\rho_{s,mn}$ represents the spatial correlation of wind speed prediction errors between wind fields located at positions $m$ and $n$. The spatial correlation matrix is symmetric. Singular value decomposition is performed on the matrix $R_S$, resulting in:

$$R_{S,h} = U_{R_{S,h}} \Sigma_{R_{S,h}} U_{R_{S,h}}^T = U_{R_{S,h}} \Sigma_{R_{S,h}}^{1/2} (U_{R_{S,h}} \Sigma_{R_{S,h}}^{1/2})^T \tag{3}$$

In the formula, $U_{R_{S,h}}$ represents the unitary matrix obtained from the composition; $\Sigma_{R_{S,h}}$ denotes represents the unitary matrix obtained from the decomposition; $U_{R_{S,h}}^T$ stands for the transpose of the unitary matrix。 The decomposed matrices are applied to $N_H$ $W_h$, forming $N_H$ partially correlated wind speed prediction error matrices $W_{S,h}$:

$$W_{S,h} = U_{R_{S,h}} \Sigma_{R_{S,h}}^{1/2} W_h \quad h = 1,2,\cdots,N_H \tag{4}$$

$N_H$ matrices $W_{S,h}$ form the spatially correlated wind speed prediction error matrix $W_S$:

$$W_S = [W_{S,1}\ W_{S,2} \cdots W_{S,h} \cdots W_{S,N_H}] \tag{5}$$

Additionally, the wind speed prediction error matrix $W_S$ needs to consider temporal autocorrelation and variance, constructing a temporal autocorrelation vector $\rho_{T,\omega}$ and a variance vector $\sigma_{T,\omega}$, as follows:

$$\rho_{T,\omega} = [\rho_{T,\omega 1}\ \rho_{T,\omega 2} \cdots \rho_{T,\omega h} \cdots \rho_{T,\omega N_H}] \tag{6}$$

$$\sigma_{T,\omega} = [\sigma_{T,\omega 1}\ \sigma_{T,\omega 2} \cdots \sigma_{T,\omega h} \cdots \sigma_{T,\omega N_H}]$$
$$\omega = 1,2,\cdots,N_W \quad h = 1,2,\cdots,N_H \tag{7}$$

$$R_{T,\omega} = \begin{bmatrix} \rho_{T,\omega 1} & \rho_{T,\omega 2} & \cdots & \rho_{T,\omega N_H} \\ \rho_{T,\omega 2} & \rho_{T,\omega 1} & \cdots & \cdots \\ \cdots & \cdots & \cdots & \cdots \\ \rho_{T,\omega N_H} & \cdots & \cdots & \rho_{T,\omega 1} \end{bmatrix} \tag{8}$$

Here, $\rho_{T,\omega h}$ represents the temporal correlation for the $\omega$th wind farm at the $h$th hour; $\sigma_{T,\omega h}$ denotes the variance; the element in the row $p$ and column $f$ of $R_{T,\omega}$ equals the $|p-f|+1$th element; the variance matrix $V_{T,\omega}$ is formed by arranging the elements of the variance vector into a diagonal matrix, as follows:

$$V_{T,\omega} = \text{diag}\{\sigma_{T,\omega}\} \tag{9}$$

Based on the matrixes $R_{T,\omega}$ and $V_{T,\omega}$, the temporal covariance matrix $K_{T,\omega}$ of the $\omega$th wind is constructed as follows:

$$K_{T,\omega} = V_{T,\omega} R_{T,\omega} V_{T,\omega} \tag{10}$$

Similar to the spatial correlation matrix, perform singular value decomposition to obtain:

$$K_{T,\omega} = U_{R_{T,\omega}} \Sigma_{R_{T,\omega}} U_{R_{T,\omega}}^T \tag{11}$$

where, $U_{R_{T,\omega}}$ represents the unitary matrix obtained from the decomposition; $\Sigma_{R_{T,\omega}}$ is the diagonal matrix; $U_{R_{T,\omega}}^T$ stands for he transpose of the unitary matrix. Let $W_{S,\omega}$ denote the row vector corresponding to the $\omega$th row elements in matrix $W_S$, The decomposed matrices are applied to the spatially correlated wind speed prediction error $W_{S,\omega}$, resulting in the spatiotemporally correlated prediction error vector $W_{ST,\omega}$:

$$W_{\text{ST},\omega} = (U_{R_{\text{T},\omega}} \Sigma^{1/2}_{R_{\text{T},\omega}} W^{\text{T}}_{\text{S},\omega})^{\text{T}} = W_{\text{S},\omega} (U_{R_{\text{T},\omega}} \Sigma^{1/2}_{R_{\text{T},\omega}})^{\text{T}} \tag{12}$$

Use the spatiotemporally correlated prediction error vector $W_{\text{ST},\omega}$ to form the spatiotemporally correlated prediction error matrix $W_{\text{ST}}(\omega, h)$:

$$W_{\text{ST}}(\omega, h) = \begin{bmatrix} W_{\text{ST},1} \\ \vdots \\ W_{\text{ST},N_{\text{W}}} \end{bmatrix} \tag{13}$$

Finally, the wind speed prediction matrix $F(\omega, h)$ is superimposed with the spatiotemporally correlated prediction error matrix $W_{\text{ST}}(\omega, h)$, resulting in the spatiotemporally correlated wind speed prediction $v(\omega, h)$.

### 3.2.2 Wind Turbine Power Output Model

The output power of a wind turbine is closely related to the wind speed and its own power curve. The model describing the wind power curve is as follows [51]:

$$P_{\text{w}}(v) = \begin{cases} 0, & v > v_{\text{o}}, v < v_{\text{i}} \\ P_{\text{r}} \dfrac{v - v_{\text{i}}}{v_{\text{r}} - v_{\text{i}}}, & v_{\text{i}} \leq v \leq v_{\text{r}} \\ P_{\text{r}}, & v_{\text{r}} < v < v_{\text{o}} \end{cases} \tag{14}$$

where, $P_{\text{w}}(v)$ represents the wind power curve; $v_{\text{i}}$ and $v_{\text{r}}$ represent the cut-in and rated wind speeds, respectively; $P_r$ represents the rated power of the wind turbine; $v_{\text{o}}$ and $v$ represent the cut-out and actual wind speeds, respectively. Based on the wind power curve, the output power of the wind turbine can be calculated. It is also assumed that reactive power compensation equipment can ensure the balance of reactive power. On this basis, the output power of the wind farm equals the sum of the output powers of all wind turbine units.

## 3 OPF-Based ATC Model

The mathematical model for the Available Transfer Capability (ATC) based on optimal power flow includes two parts: the objective function and constraints. The objective function is the maximum transferable electric power capacity across inter-regional transmission interfaces; the model's equality constraints include not only the power flow equations under normal operating conditions but also generalized parametric power flow equations considering load margins, and incorporates the spatiotemporally correlated wind power output into the power flow equations. The inequality constraints include the upper and lower bounds of active power output from generators, the upper and lower bounds of reactive power output, line flow constraints, and node voltage boundaries. The mathematical model for ATC can be described as:

$$\max \quad f \\ s.t. \begin{cases} g(x)=0 \\ \underline{h} \le h(x) \le \overline{h} \end{cases} \tag{15}$$

The objective function is as below：

$$f = [\sum_{i \in A, j \in B}(P_{i,j} - P_{i,j}^0)] \tag{16}$$

In the formula, , $f$ represents the objective function value, which is the power that can further be transmitted across the transmission interface; $j$;$P_{i,j}$ represents the active power transmitted from node $i$ to $j$, which is a function of the voltages and phase angles at various nodes; 0 represents the base state value, which is a constant; $A$ and $B$ represent the sets of nodes in the sending and receiving areas, respectively.

The mathematical expression for the equality constraints $g(x)=0$ is as follows：

$$P_{wi} + P_{Gi} - bP_{Di} = U_i \sum_{j=1}^{n} U_j (G_{ij} \cos\theta_{ij} + B_{ij} \sin\theta_{ij}) \\ Q_{wi} + Q_{Gi} - bQ_{Di} = U_i \sum_{j=1}^{n} U_j (G_{ij} \sin\theta_{ij} - B_{ij} \cos\theta_{ij}) \tag{17}$$

In the formula, $P_{wi}$ and $Q_{wi}$ represent the active and reactive power injected by wind power at node $i$, where $P_{wi}=P_w(v(\omega, h))$ and $v(\omega, h)=F(\omega, h)+W_{ST}(\omega, h)$; $P_{Gi}$ and $Q_G$ represent the active and reactive power output of conventional generator $i$; $P_{Di}$ and $Q_{Di}$ represent the active and reactive loads at node iii; $U_i$ and $U_j$ represent the voltage magnitudes at nodes $i$ and $j$; $\theta_{ij}$ represents the phase angle difference between nodes $i$ and $j$, where $\theta_{ij}=\theta_i-\theta_j$ ; $b$ represents the load fluctuation coefficient; $G_{ij}$ and $B_{ij}$ represent the real and imaginary parts of the admittance matrix elements in the $i$th row and $j$th column. This model uses the polar form representation of node voltages.

The mathematical expressions for the inequality constraints are as follows [52]：：

$$\begin{aligned} P_{Gi,\min} &\le P_{Gi} \le P_{Gi,\max}, & i \in S_G \\ Q_{Gi,\min} &\le Q_{Gi} \le Q_{Gi,\max}, & i \in S_G \\ P_{L,\min} &\le P_L \le P_{L,\max}, & L \in S_L \\ U_{i,\min} &\le U_i \le U_{i,\max}, & i \in S_N \end{aligned} \tag{18}$$

where, $P_{Gi,\max}$ and $P_{Gi,\min}$ represent the upper and lower bounds of the active power output of the generator;; $Q_{Gi,\max}$、$Q_{Gi,\min}$ are the upper and lower bounds of the reactive power output of the generator; $P_L=[U_iU_j(G_{ij}\cos\theta_{ij}+B_{ij}\sin\theta_{ij})-U_i^2 G_{ij}]$, where $P_L$ represents the active power flowing through line L;$P_{L,\max}$、$P_{L,\min}$ stand for the upper and lower bounds of the active power flowing through line L; $U_{i,\max}$ and $U_{i,\min}$ refer to the upper and lower voltage bounds at node $i$; $S_G$, $S_L$ and $S_N$ are the sets of all generators, lines, and nodes, respectively.

# 4 ATC Calculation Process

This paper considers the effects of spatiotemporal correlation of wind speeds and load fluctuations on the system, incorporating wind power output with spatiotemporal correlation and load fluctuations into the power flow equations to establish an ATC calculation model based on Optimal Power Flow (OPF). The modified model is optimized using the Primal-Dual Interior Point Method (PDIPM). By introducing slack variables, inequality constraints are converted into equality constraints, transforming the objective function into a barrier function form. A Lagrangian function is constructed, and the nonlinear system of equations is solved using the Newton-Raphson method to determine the optimal search direction for each variable. The variables are adjusted in each iteration, and when the complementarity gap meets the specified accuracy, the optimal solution is output. The ATC calculation steps are as Follows:

Step 1: First, initial wind speed predictions are obtained using meteorological simulation tools, and the random errors in wind speed prediction are represented using a matrix of independently and identically distributed Gaussian random variables [53]. Next, spatial correlation matrices and temporal covariance matrices are constructed and subjected to singular value decomposition. The resulting unitary and diagonal matrices are then applied to the wind speed prediction error matrix, producing a final wind speed prediction error matrix with spatiotemporal correlation.

Step 2: The wind speed prediction errors with spatiotemporal correlation are superimposed with the original wind speed predictions to obtain more accurate wind speed predictions. These wind speeds are then used to calculate wind power output, achieved by applying the power curve model of the wind turbines.

Step 3: The calculated wind power output with spatiotemporal correlation and data considering load fluctuations are integrated into the power flow equations to establish an ATC calculation model based on optimal power flow [54-55].

Step 4: The aforementioned model is applied under various grid operation scenarios and optimized using the Primal-Dual Interior Point Method, comparing changes in ATC across different scenarios to derive detailed research conclusions.

Through this series of steps, this study aims to provide a more accurate and practical tool for assessing and optimizing the impact of wind power integration on power systems, thereby supporting decision-making and operations in the power market.

## 5 Case Study Analysis

To verify the effectiveness of the ATC calculation model, and to discuss the impact of different scenarios on ATC, wind farm simulations are conducted based on the modified IEEE-39 bus system [56-57]. The system structure diagram is shown in Fig. 1.

The system consists of 10 generators and 46 transmission lines, with a total load of 6254.23 MW, divided into two parts: a sending area and a receiving area, interconnected by tie lines 14-15, 17-18, and 25-26. This paper conducts case studies using examples of two wind farms and multiple wind farms. Each wind turbine has a rated power of 5 MW, a cut-in wind speed of 3 m/s, a rated wind speed of 13 m/s, and a cut-out wind speed of 25 m/s. The standard deviation of the wind speed prediction error is 0.05, and the mean value of the wind speed prediction error is 0.

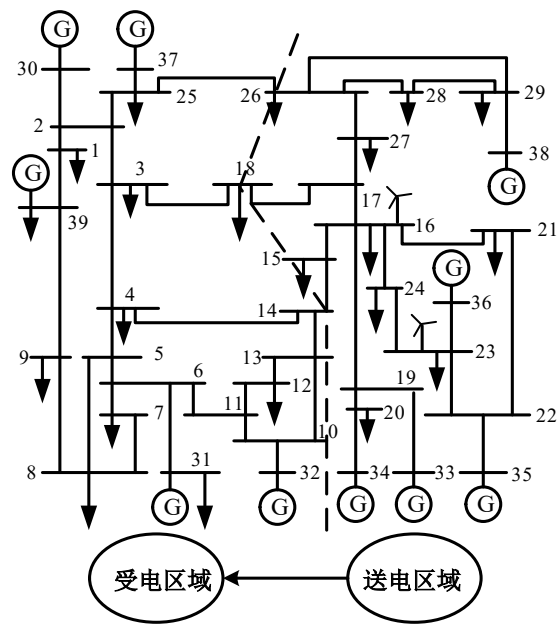

**Fig. 1. Structure Diagram of the Modified IEEE-39 Bus System**

Each wind farm consists of 60 wind turbines, each with a rated output power of 5 MW, giving the wind farm an installed capacity of 300 MW [58-60]. The two wind farms are connected to nodes 16 and 23, respectively, as shown in Fig. 1. The analysis is as follows:

## 5.1 Impact of Interconnecting Two Wind Farms on System ATC

(1) Wind Farm Simulation Considering Spatiotemporal Correlation

The spatial correlation matrix for the two wind farms is as follows:

$$R_S = \begin{bmatrix} 1 & \rho \\ \rho & 1 \end{bmatrix} \quad (19)$$

The temporal autocorrelation distribution function of the wind speed series is shown in Fig. 2:

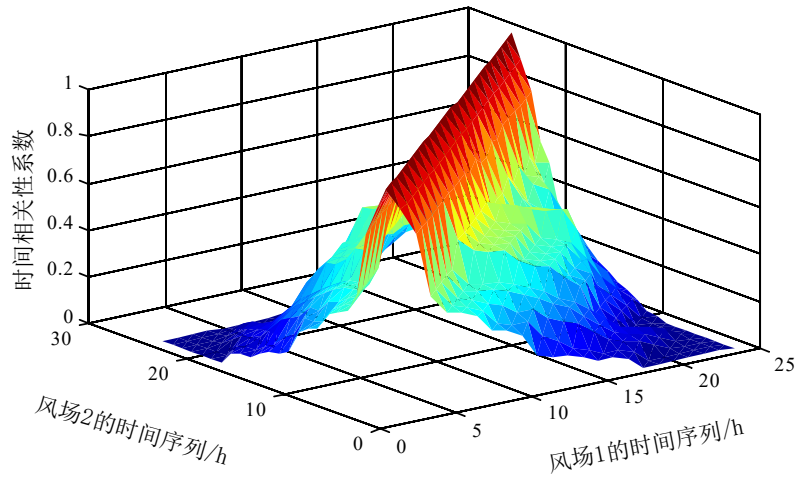

**Fig. 2. Autocorrelation Function of the Wind Speed Series**

Assuming no spatiotemporal correlation, the wind speeds at the two wind farms are the same over a 24-hour period. After considering spatiotemporal correlation, when the correlation coefficient $\rho$ is 0.5, the wind speed curves for the two wind farms are shown in Fig. 3. The wind speed curves for Wind Farm 1 under different correlation coefficients, obtained through simulation, are shown in Fig. 4:

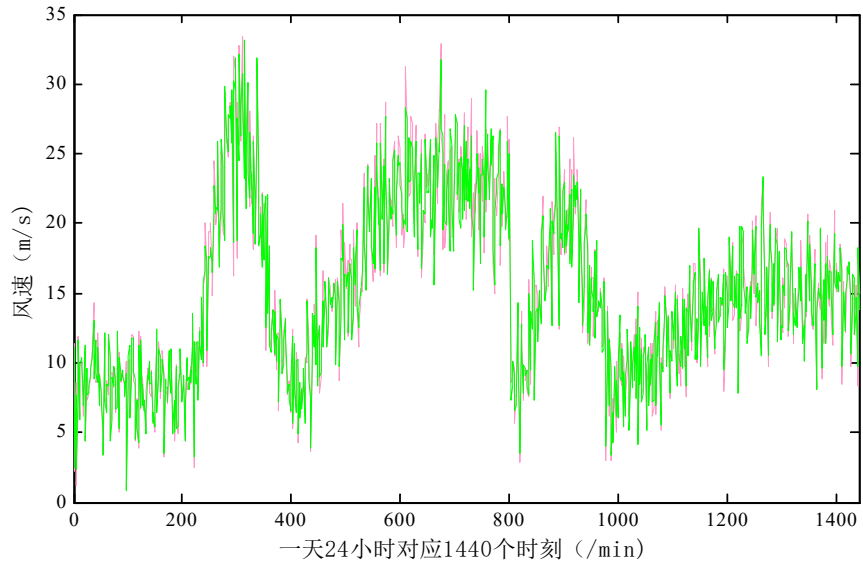

**Fig. 3. Wind Speeds of the Two Wind Farms**

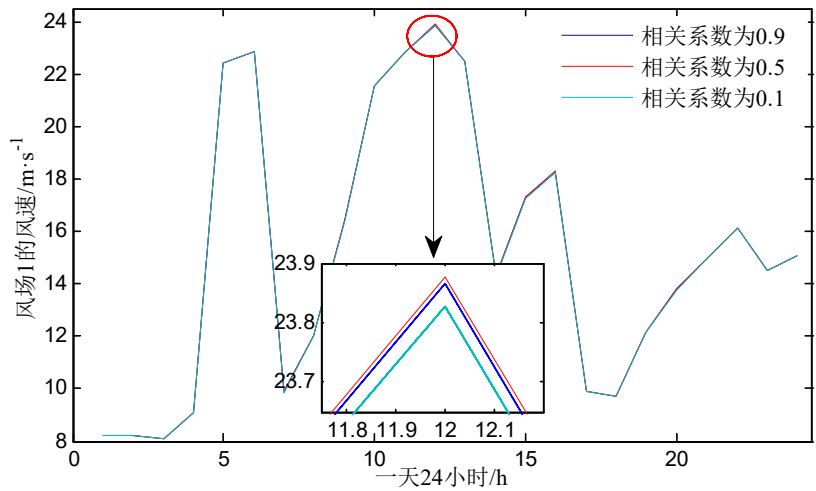

**Fig. 4. Wind Speed Curves for Wind Farm 1 under Different Correlation Coefficients**

From Figs. 3 and 4, it can be seen that the correlation coefficient of wind speed has a certain impact on the predicted values of the wind farms. For the same wind farm, the correlation coefficient of wind speed has different impacts at different times, and the impact is greatest when the wind speed reaches its maximum value. At the peak moment, when the correlation coefficient is 0.5, the impact on wind speed values is the greatest.

**(2) Analysis of System ATC after Wind Power Integration**

Depending on the conditions of wind power integration, large wind farms have different impacts on ATC. Considering the spatiotemporal correlation of wind speeds and load fluctuations, this section analyzes the impact of wind speed correlation coefficients, wind power integration capacity, wind farm integration locations, and integration methods on ATC.

First, the impact of wind speed correlation on ATC is analyzed. Wind farms with a capacity of 300 MW each are added to nodes 16 and 23 on the sending side. The impact of different correlation coefficients $\rho$ on system ATC is discussed, and the simulation results are shown in Fig. 5.

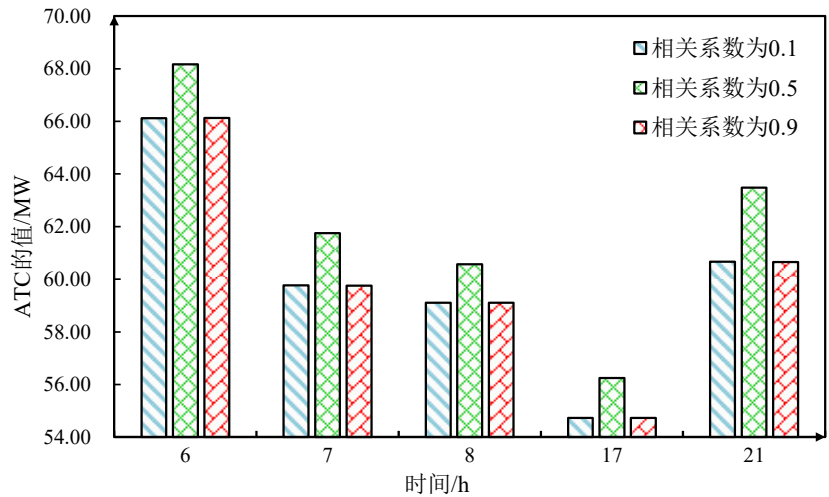

**Fig. 5. Impact of Wind Speed Correlation on ATC**

From Fig. 5, it can be seen that different correlation coefficients have varying impacts on ATC. At the same moment, as the correlation coefficient increases, ATC first increases and then decreases, with the correlation coefficient of 0.5 having the greatest impact on the system's ATC. Additionally, at different times, the correlation coefficient exhibits the same trend in its impact on ATC. From this, it can be concluded that as the correlation coefficient increases, the system's ATC first decreases and then increases. This is because different correlation coefficients affect the output power of the wind farms, altering the system's power flow distribution, which in turn causes changes in the system's ATC.

Next, the impact of wind power integration capacity on ATC is analyzed, with the results shown in Fig. 6.

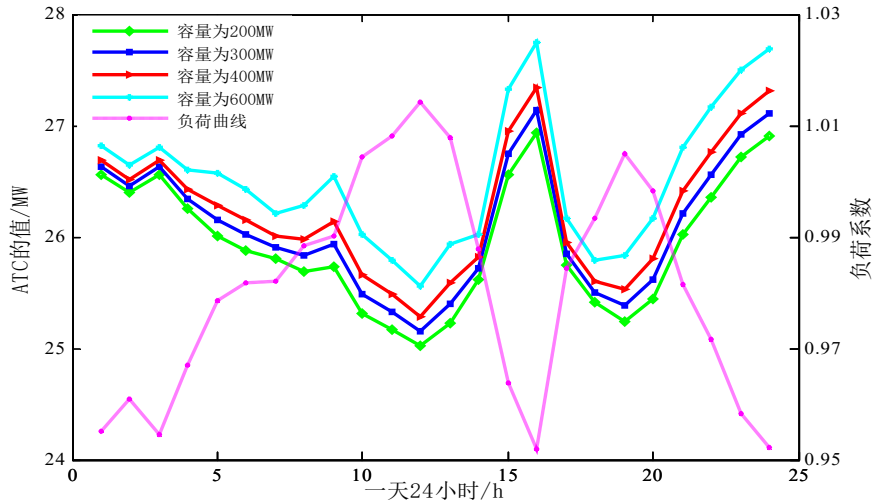

**Fig. 6. Fluctuation Curve of ATC with Wind Farms of Different Capacities Connected**

From Fig. 6, it can be seen that when wind farms of different capacities are connected to the system, the trend in ATC changes is essentially the same and shows an opposite trend to that of load fluctuations. This is because the temporal changes in ATC are mainly caused by the temporal changes in load; within a certain range, as the capacity of wind power integration increases, the value of ATC also increases. Some statistical data indicate that the limit of wind power integration capacity generally does not exceed 10% of the total system load. The simulations in this paper show that when the wind power integration capacity reaches 800 MW, the value of ATC no longer increases. This is due to the fixed capacity of the system to accept wind power; when the line power reaches its upper limit, the transmission power across the transmission interface no longer increases, and the value of ATC remains unchanged.

The ATC fluctuation curve, simulated every 15 minutes over a 24-hour period, is shown in Fig. 7.

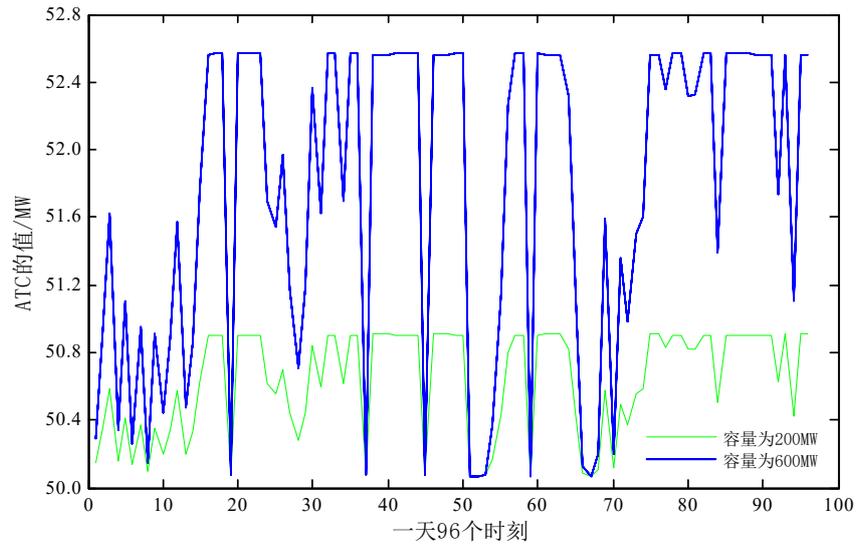

**Fig. 7. Fluctuation Curve of ATC**

Finally, the impact of wind farm connection locations and methods on ATC is discussed:

1) Discussing the impact of wind farm grid connection locations on ATC, select specific moments for ATC analysis, with results as follows:

**Table 1 Impact of Different Wind Farm Connection Locations on ATC**

| Wind Farm Connection Location | | Time /h | | | |
|---|---|---|---|---|---|
| | | 3 | 6 | 15 | 22 |
| No wind power | | 109.583 | 59.544 | 93.374 | 79.612 |
| Receiving side | (16, 4) | 110.062 | 60.836 | 94.923 | 81.250 |
| | (16, 6) | 110.058 | 60.829 | 94.910 | 81.236 |
| | (16, 8) | 110.059 | 60.832 | 94.914 | 81.240 |
| | (16, 25) | 110.059 | 60.818 | 94.914 | 81.242 |
| | (16, 31) | 110.058 | 60.830 | 94.910 | 81.236 |
| Sending Side | (16, 15) | 110.073 | 60.858 | 94.959 | 81.288 |
| | (16, 19) | 110.082 | 60.876 | 94.991 | 81.324 |
| | (16, 21) | 110.078 | 60.866 | 94.976 | 81.307 |
| | (16, 22) | 110.078 | 60.868 | 94.979 | 81.311 |
| | (16, 23) | 110.077 | 60.866 | 94.976 | 81.306 |
| | (16, 35) | 110.078 | 60.869 | 94.981 | 81.333 |

In Table 1, nodes 4, 6, 8, 25, and 31 are located on the receiving side, while nodes 15, 19, 21, 22, 23, and 35 are on the sending side. By comparing scenarios with and without wind power integration, it can be observed that after wind power is connected to the system, ATC increases in all cases. This indicates that even adding a fluctuating power source like a wind farm can enhance the system's ATC. When both wind farms are connected on the sending side versus both the sending and receiving sides, the former shows a higher ATC. Therefore, connecting wind power on the sending side is beneficial for improving the system's ATC.

2) Impact of wind power integration methods on system ATC

a) Two wind farms with a capacity of 100 MW each are connected at node 38 on the sending side, replacing the output of traditional generating units originally totaling 200 MW. b) Two wind farms with a capacity of 100 MW each are connected at node 39 on the receiving side, replacing the output of traditional generating units originally 200 MW.

**Table 2 Impact of Wind Power Integration Methods on ATC**

| Integration Method | Time/h | | | | |
|---|---|---|---|---|---|
| | 1 | 4 | 10 | 14 | 19 |
| a | 129.360 | 100.364 | 19.946 | 47.441 | 19.208 |
| b | 130.029 | 101.070 | 19.963 | 47.454 | 19.222 |

Comparing scenarios a and b, it is evident that integrating wind power in the sending area to replace conventional units reduces ATC, whereas integrating wind power in the receiving area to replace conventional unit output increases ATC. This is because wind power output is fluctuating and random, and cannot adapt to changes in the system's power flows like conventional units can. Replacing conventional units with wind power during optimization reduces the optimization range, thereby decreasing the system's ATC value. Therefore, when planning the system, it is preferable to integrate wind farms on the receiving side to replace traditional units [61]. This not only reduces air pollution caused by the output of traditional units but also enhances the system's ATC, thus helping to promote the safe and stable operation of the system [62].

## 5.2 Analysis of ATC with Multiple Wind Farms Integrated

Taking four wind farms as an example, each farm has 20 wind turbines with a rated output power of 5 MW and an installed capacity of 100 MW, integrated at nodes 3, 4, 16, and 23. The impact of wind speed correlation coefficients, wind farm capacity, integration locations, and integration methods on ATC has already been discussed in detail in the case studies involving two wind farms. Here, we will only discuss the impact of wind speed correlation coefficients on ATC, with the analysis results in Fig. 8.

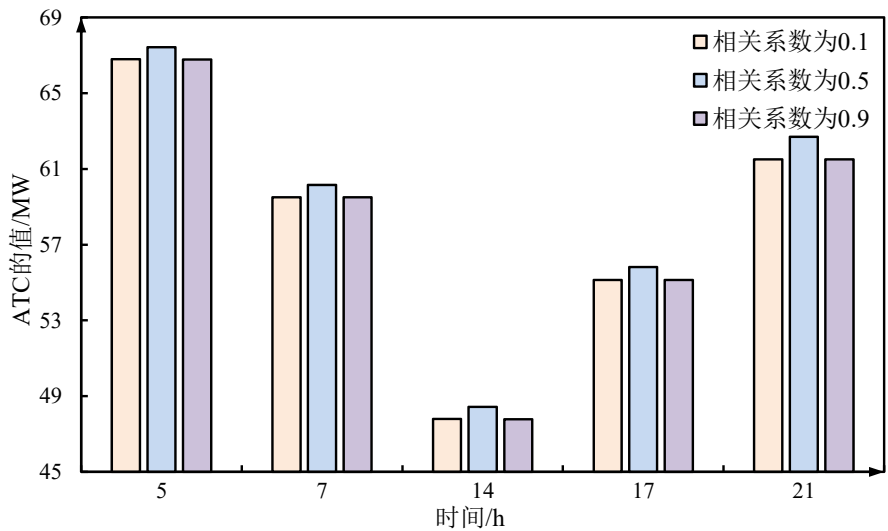

Fig. 8 Impact of Wind Speed Correlation on ATC

From Fig. 8, it can be seen that as the correlation coefficient increases, ATC initially increases and then decreases, with the greatest impact observed at a correlation coefficient of 0.5. Therefore, when planning wind farms, it is important to consider the impact of wind speed spatiotemporal correlation on ATC. Comparing the case studies of two wind farm systems and multiple wind farm systems, it is evident that wind speed correlation affects both systems in the same way, demonstrating the accuracy of the constructed model [63-64].

## 6 Conclusion

This study addresses the complexities of wind power integration into power systems, considering the spatiotemporal characteristics of wind speed and load fluctuations, and has successfully developed an ATC calculation model based on optimal power flow. By incorporating the Primal-Dual Interior Point Method (PDIPM), this model not only improves the accuracy and efficiency of calculations but also enhances the ability to handle the nonlinearity and complexity of the model, significantly boosting performance in solving problems. Methodologically, this paper innovatively constructs a wind speed prediction error matrix that accounts for spatiotemporal correlation. This technique, by integrating wind speed prediction errors with actual measurements, generates highly accurate spatiotemporal correlated wind speed predictions, providing key inputs for precise ATC calculations. Additionally, this method effectively addresses the randomness and intermittency of wind power output, enhancing the prediction accuracy of the impacts of wind power integration. By applying this model in various grid scenarios, this research has detailed the effects of wind speed correlation coefficients, wind power integration capacity, integration locations, and integration methods on ATC.

These analyses not only validate the effectiveness and applicability of the model but also clarify how these variables influence ATC values, offering valuable decision support for risk management in power systems, wind farm planning, and power market operations. The results of this paper not only provide a powerful computational tool for the stability and safe operation of power systems but also offer theoretical foundations and practical guidance for the further development and optimization of wind power technology.